# Theory assessment and reality in Boltzmann's epistemological thinking

*Marcelo B. Ribeiro*[1,2]
*and*
*Antonio A. P. Videira*[3,4,5]

This paper discusses how theories can be assessed within the epistemological viewpoint advanced by the Austrian physicist Ludwig E. Boltzmann. It builds upon, and further develops, the perspective of Boltzmann's thinking as advanced by Ribeiro and Videira (1998). Boltzmann's epistemological viewpoint accepts that reality is real and proposes that reality can be described by different points of view because his main philosophical thesis states that scientific theories are images of Nature. We present the historical context that witnessed the genesis of Boltzmann's ideas and expand Ribeiro and Videira's (1998) perspective by arguing that later in his life Boltzmann realized the insufficiency of his thesis as justification for theoretical pluralism and avoidance of dogmatism. Consequently, his thinking went beyond epistemology, the nature of scientific knowledge, to include realism, the nature of the represented objects.[6]

# 1. Introduction

The second half of the nineteenth century witnessed heated debates about the possible existence of the atom, and even the need for such a concept in physics. Contributors to these debates included some leading natural scientists

1 Physics Institute, Universidade Federal do Rio de Janeiro, Rio de Janeiro, Brazil; E-mail: mbr@if.ufrj.br; ORCID 0000-0002-6919-2624

2 Graduate Program in Applied Physics, Universidade Federal do Rio de Janeiro, Rio de Janeiro, Brazil

3 Department of Philosophy, Universidade do Estado do Rio de Janeiro, Rio de Janeiro, Brazil; E-mail: guto@cbpf.br; ORCID 0000-0003-4369-9221

4 Graduate Program in History and Teaching of Mathematics and Physics, Universidade Federal do Rio de Janeiro, Rio de Janeiro, Brazil

5 Center for Humanistic and Philosophical Studies, Universidade Católica Portuguesa, Braga, Portugal

6 The authors would also like to thank C. Roy Keys Inc. for allowing us to use previously published materials in a revised and expanded form in this article. Most of the Sections 2 and 3 in this article first appeared in Ribeiro and Videira (1998).

such as Ludwig Boltzmann (1844-1906), Ernst Mach (1838-1916), Wilhelm Ostwald (1853-1932), and Georg Helm (1851-1923). These debates attracted the attention of other scientists, including Arnold Sommerfeld (1868-1951), and later became the subject of analyses and discussions among historians and philosophers of science. Experts on the history and philosophy of science became particularly interested in the viewpoints professed by this group of scientists towards atomism. In this context, Boltzmann is generally described as a scientist who was inclined to accept not only the existence of, but also the need for, an atomic conception of the physical world. Considering the fact that Boltzmann was an atomist, which would also make him a realist, doubts remain to this day about what kind of realism underpinned his scientific positions.

This paper offers arguments in favor of the thesis that Boltzmann was a realist, while also presenting some further consequences of his epistemological thinking. Boltzmann in fact developed a very particular form of realism because he rejected the notion that a single type of realism could be valid for all sciences. In other words, his realism can be seen as weak or, at most, moderate. His approach to realism can also be observed in the stance he adopted in public debates and conferences and in his academic papers.

Boltzmann's academic activities began in 1863, when he enrolled at the University of Vienna to study natural sciences, completing his doctorate in physics in 1866, when he also published a paper on the mechanical theory of heat (Darrigol 2018). This choice of topic clearly demonstrates his early interest in the fundamentals of physics – an interest he maintained throughout his career. He held teaching, research, and administrative positions in Vienna, followed by Graz, Munich, back to Vienna, Leipzig, and again Vienna, where he took up his final position in 1902 and which he retained until his death in 1906. Boltzmann's willingness to work at different universities was directly related to his most basic ambition: to contribute to the progress of science and to the creation and consolidation of theoretical physics as an autonomous discipline.

Boltzmann's contributions to the philosophy of science were born out of his intense intellectual exchanges with many other eminent scientists of his time, like Hermann von Helmholtz (1821-1924), Heinrich Rudolf Hertz (1857-1894), Pierre Duhem (1861-1916), Henri Poincaré (1854-1912), and Max Planck (1858-1947), as well as the aforementioned Ostwald, Helm, and Mach, on the aims and methods of theoretical physics. The intensity and duration of his participation in these debates show how important he felt they were for determining the nature of theoretical physics. He did not seek to establish a philosophy for physics, much less a philosophy of science in general; his epistemological style was to defend his views on the fundamentals of science against the criticisms of other scientists and philosophers, who, in his opinion, were mistaken as to the nature of scientific theories.

The way Boltzmann presented his epistemological viewpoints was also due to one of the epistemological principles he held most dear, namely, that in the process of gaining hegemony among the members of a scientific community, no scientific theory could, for this very reason, exclude other theories. According to Boltzmann, the exclusion of other theories would block progress in science, since such dogmatic behavior would lead to nothing positive and would end up impoverishing the scientific enterprise. That is why, when participating in scholarly debates, he championed an open attitude towards the analysis of various theories. His pro-pluralism stance was reinforced by his belief that science would, in a not-too-distant future, be deeply transformed in a way that nobody at his time could foresee with any degree of certainty.

Boltzmann viewed the end of the nineteenth century as a time of doubts, mistrust, and growing skepticism on the part of scientists vis-a-vis the progress of science, Many were convinced that major changes were inevitable, making them more inclined to defend their ideas in the hope that they might help lift science out of its theoretical stalemate. This impasse was due to the inconclusiveness of the debates about the importance of hypotheses in physics, how a theory is built upon, whether empirical known facts must always be the starting point, whether scientific ingenuity and creativity could be used freely, and even whether physical theories should describe natural phenomena rather than explain them, putting aside the old ideal of identifying their real causes for good.

Boltzmann's philosophical analyses are rich, complex, and even contradictory, if taken to their limits. In other words, they might be seen as containing an internal tension which, if they were applied inappropriately, would lead to the conclusion that his thinking was unsustainable because it was contradictory. Such a conclusion only holds if it is felt that the philosophizing of professional physicists ought to follow the same evaluation criteria, coherence, and rigor as sought by professional philosophers. Considering Boltzmann's standpoint regarding professional philosophers, this conclusion is not valid.[7] The answer to the question as to why Boltzmann's thinking could be seen as contradictory lies in how he viewed the philosophical positions advanced by natural scientists at the end of the nineteenth century. Below we summarize his perspective on this debate.

The physicists who participated in philosophical debates on the natural sciences at the end of the nineteenth century can be divided into two camps.

7 See Videira (1992, 139-142) for more details.

The first advocated the thesis that scientific theories could only describe natural phenomena, whereas the second claimed that this viewpoint was naive because any description is to some extent a construct.

The first camp was known as *descriptive*, or *positivist*,[8] in that its followers deliberately sought to avoid confusing any description with what was being described, presupposing that empirical data were the only source material that could be employed to advance a description. By using sensory data, they promoted what they argued was the universality of scientific knowledge, which they felt would bring about a healthy reduction in the degree of arbitrariness of the concepts employed in the formulation of theories and models in physics. This viewpoint is generally seen as anti-realistic because it leads to the notion that science cannot be mistaken for what is real, since what is real cannot be accessed by scientists. Strictly speaking, belief in the existence of an independent, external real world is held as detrimental, as it may encourage us to seek something that cannot be sought through our senses and reason. The human senses are therefore incapable of attaining the essence of natural phenomena or determining their causes. In other words, with our senses we can perceive the behavior of natural phenomena, but not why they behave in one way and not another. We cannot know if natural phenomena could behave differently than the way we perceive them to behave. Hence, at the basis of this position lies the belief that we can only "touch" what is real and only "walk" on its surface.

The second philosophical camp supported by physicists at the end of the nineteenth century was known as *representationism*.[9] Its followers opposed the positivists because they explicitly stated that our senses are not enough of themselves to fulfill the task of formulating models and theories. Concepts need to be created by scientists upon foundational pillars, which may include data obtained by our senses. Nevertheless, these pillars cannot be reduced to the data, since preferences of all types – methodological, epistemological, ontological, aesthetic, etc. – can be rightfully employed during the formulation of what is subsequently advanced as representing Nature. There were also concerns regarding the degree of arbitrariness in such representations, and hence it was advisable to compare representation with Nature as soon as possible. Supporters of this representative viewpoint stated that simple comparison was not enough to reduce this degree of arbitrariness. In fact, they argued that there would always be some arbitrariness, which could only be overcome by adopting theoretical pluralism. And a plurality of theories could only work by means of critical dialogue.

8 Boltzmann sometimes used the term *phenomenalist*.

9 Representationism should not be confused with the different term *representationalism*.

According to the reasoning above, it seems very plausible to consider the descriptive viewpoint as being subsumed by the representationist viewpoint. However, *representationism* cannot be confused with *descriptivism* in the sense that the latter may lead to *solipsism*, considered an absurd position by physicists in general and Boltzmann in particular. The only way to avoid solipsism would be to adopt *realism*, which presupposes the existence of an externally independent real world, or Nature.

Boltzmann was aware that these two camps coexisted in the natural sciences in a manner that was not always peaceable or amicable. The two camps were mutually exclusive as their proponents each saw theirs as the only viewpoint serious scientists could possibly adopt. Any attempts to reconcile them would lead to a painful contradiction. On occasion, Boltzmann's writings exhibit a certain tendency towards the descriptive position in terms of its intelligibility, but ultimately he viewed descriptivism as a particular case of his own perspective on representationism. Boltzmann sometimes appears as a positivist, and other times as a realist.

Seeing descriptivism as a particular case of representationism can only work if it is understood in epistemological terms, although this maneuver is less tenable ontologically. In philosophical and ontological terms, the idea that a physicist might simultaneously defend positivism and realism is unusual at the least. However, as Boltzmann saw it, theoretical pluralism was a key to enabling these two rival philosophical positions to coexist, not because this could be harnessed to build valid scientific solutions, but because it constituted a realization of his most cherished principle, namely, that scientific progress can only be achieved through the interaction, coexistence, and competition not only of different, but even of rival scientific theories and models. We return to this point below.

When participating in these epistemological discussions, Boltzmann sought firstly to assure the survival of the theories he believed in, while also guaranteeing a place for others. The ability of a certain theory to predict new phenomena did *not* mean, he argued, that it can predict its own future, much less the future of science. On the other hand, a theory that had already produced good results should not be abandoned. Recognizing the scientific limits of a theory did not mean it should be banished from the realm of science. The incapacity of a theory to predict its own future was probably the main factor that motivated Boltzmann to try to better understand the process and progress of science. His interpretation of Darwinism served as a basis for him to be able to reach some of the conclusions he reached, which afterwards paved the way for him to take the path he intended to follow. For Boltzmann a scientific theory is nothing more than a *representation* of Nature.

In this article we argue that Boltzmann realized along the years that the defense that scientific theories are nothing more than representations of natural phenomena was not enough to justify theoretical pluralism and, as a result, avoid dogmatism. For that it was necessary to move the discussion from the nature of scientific knowledge to the nature of the objects present in the representations. In other words, he recognized that epistemology is insufficient to sustain his defense of pluralism. Hence, this article expands the discussion from theories being images of Nature, which was the focus of Ribeiro and Videira (1998), to *Boltzmann's concept of reality*.

The plan of the paper is as follows. Section 2 reviews Boltzmann's concepts of realism and scientific truth as discussed in Ribeiro and Videira (1998). Section 3 applies the concepts of the previous section to theory assessment. Section 4 compares the Boltzmann's perspective discussed in the previous sections with some of his modern interpreters on Boltzmann's realism. Section 5 presents our conclusions.

# 2. Models, realism, and scientific truth

The question of the existence of the external world, or matter, must be seen in the light of another problem: whether the answer to this question complicates or simplifies our image of the world (*weltbild*) (Boltzmann 1905a). Boltzmann seems to feel the need to avoid pointless discussions of the type often engaged in by philosophers. Even though he recognized that there was no definitive evidence to support or refute the existence of matter, at least in his time, he felt that belief in either position was a matter of ideology. Although he did not define what he meant by ideology, it seems that for him this word had a negative connotation. Both idealism and realism were ideologies. Another example of Boltzmann’s views on ideology – and arguably more important than the previous one, but nonetheless contentious – has to do with solipsism. Boltzmann was uncompromisingly anti-idealism, since idealism denied the existence of the external material world.

Boltzmann seems to have believed that it is impossible not to choose between idealism and realism. Insofar as it is impossible to prove which position is best, choosing between them can only be done by evaluating the weak points of each position. Evaluating the weaknesses of idealism, which could not overcome the gap between what is alive and what is not alive, Boltzmann opted for realism (Rosa et al. 2020). Boltzmann was careful in the way he referred to these philosophical systems, stating that they were solely modes of expression (*Ausdruckweise*) used by scientists and philosophers to convey their ideas about reality. Choosing the most appropriate or most adequate mode would do away with false problems. Less time would be spent trying to answer false questions, which, according to Boltzmann, was a major obstacle to be avoided. This choice of the most adequate way to express science leads us directly to the core of the Austrian physicist's epistemological thinking.

Boltzmann sought to show at the end of the nineteenth century that all scientific theories are nothing more than representations, that is, constructions of natural phenomena. As representations, scientific theories cannot aim to know Nature itself, since such ultimate knowledge is, and will ever be, unknowable. A scientific theory will never be completely or definitively true. This viewpoint actually redefines the concept of scientific truth by advancing the notion that the identification of a theory with researched objects is a *weak* one; i.e., such identification *(1)* cannot be unique, *(2)* cannot be complete, and *(3)* is temporarily limited, since scientific theories are nothing more than images of Nature. As a consequence, *(1.1)* the same aspects of Nature [10] can be represented by more than one, often competing, theory for the preference of the scientific community, *(2.1)* as representations, scientific theories will never be able to show all aspects of natural phenomena, inasmuch as such complete knowledge is unknowable, and *(3.1)* any scientific theory can one day be replaced by another. It is the potential for one theory to be replaced by another that defines and constitutes scientific progress.

Boltzmann's ideas about scientific models as representations are clearly stated when he examines the notion of model.

> Models in the mathematical, physical and mechanical sciences are of the greatest importance. Long ago philosophy perceived the essence of our process of thought to lie in the fact that we attach to the various real objects around us particular physical attributes – our concepts – and by means of these try to represent the objects to our minds. Such views were formerly regarded by mathematicians and physicists as nothing more than unfertile speculations, but in more recent times they have been brought by J. C. Maxwell, H. v. Helmholtz, E. Mach, H. Hertz and many others into intimate relation with the whole body of mathematical and physical theory. On this view our thoughts stand to things in the same relation as models to the objects they represent. The essence of the process is the attachment of one concept having a definite content to each thing, but without implying complete similarity between thing and thought; for naturally we can know but little of the resemblance of our thoughts to the things to which we attach them. What resemblance there is, lies principally in the nature of the connexion [sic], the correlation being analogous to that which obtains between thought and language, language and writing. (…) Here, of course, the symbolization of the thing is the important point, though, where feasible, the utmost possible correspondence is sought between the two (…) we are simply extending and continuing the principle by means of which we comprehend objects in thought and represent them in language or writing (Boltzmann 1974a, 213).

10 Here the words "Nature" and "external world" are used as synonyms.

The most important epistemological conclusion Boltzmann reached, and one that lies at the heart of his philosophical thinking, is usually called *theoretical pluralism*. This is a consequence of the thesis that all scientific theories are representations of Nature. If a scientific theory is a representation, then it begins life as a free creation of the scientist, who can formulate it from a purely personal perspective, selecting which metaphysical presuppositions, theoretical options, and mathematical language to use, and even dismissing some observational data. This is valid when the theory is being formulated. However, once formulated, for a theory to be eligible to become part of science, it must be confronted by experience. If it does not pass this crucial test, the theory must be reformulated or even put aside. Boltzmann also stressed that since all scientific theories are, to some extent, the free creations of scientists, scientific work is impossible without the use of theoretical concepts, which originate from the fact that no scientific theory can be formulated from the mere observation of natural phenomena.

Theoretical pluralism also states that the same natural phenomenon can be explained by different theories. This stems from the fact that, as discussed above, any theory is a representation, a construction, an image of the natural external world, and nothing more. In Boltzmann's view, this was the only way science could be done. Theories were only scientific if they were a construction, a representation. In his words:

> (…) Hertz makes physicists properly aware of something philosophers had no doubt long since stated, namely that no theory can be objective, actually coinciding with nature, but rather that each theory is only a mental picture of phenomena, related to them as sign is to designatum.
>
> (…) From this it follows that it cannot be our task to find an absolutely correct theory but rather a picture that is as simple as possible and that represents phenomena as accurately as possible. One might even conceive of two quite different theories both equally simple and equally congruent with phenomena, which therefore in spite of their difference are equally correct. The assertion that a given theory is the only correct one can only express our subjective conviction that there could not be another equally simple and fitting image (Boltzmann 1974b, 90-91).

In summary, theoretical pluralism synthesizes the idea that as knowledge of Nature itself is impossible, each theory can only be better, or worse, than other theories. This is the mechanism that prevents science from stagnating. Within this perspective, truth can only be *provisional*, and is in fact an approximation achieved by different theoretical images.

An important corollary of Boltzmann's theoretical pluralism is his notion of scientific truth. Since Boltzmann's main thesis states that all scientific theories are representations of natural phenomena the concept of truth in modern science should no longer be about determining Nature itself. Therefore, such strong correspondence is not valid, and should be replaced by a *weak correspondence*, which in turn enables scientists to choose one model among several possible ones, since more than one model, or theory, may well represent the same group of natural phenomena and/or experimental data. At this moment Boltzmann advances another definition of scientific truth: *adequacy*: theory A is more adequate than theory B if the former is capable of explaining more intelligibly a certain set of natural phenomena than the latter.

> (…) let me choose as goal of the present talk not just kinetic molecular theory but a largely specialized branch of it. Far from wishing to deny that this contains hypothetical elements, I must declare that branch to be a picture that boldly transcends pure facts of observation, and yet I regard it as not unworthy of discussion at this point; a measure of my confidence in the utility of the hypotheses as soon as they throw new light on certain peculiar features of the observed facts, representing their interrelation with a clarity unattainable by other means. Of course we shall always have to remember that we are dealing with hypotheses capable and needful of constant further development and to be abandoned only when all the relations they represent can be understood even more clearly in some other way.
>
> (…) We must not aspire to derive nature from our concepts, but must adapt the latter to the former. We must not think that everything can be arranged according to our categories or that there is such a thing as a most perfect arrangement: it will only ever be a variable one, merely adapted to current needs (Boltzmann 1974c, 163, 166).

He also noted that since theories are images of Nature, they all have some explanatory power, and that a good theory is one that is carefully crafted by scientists in a process similar to Darwin's natural selection (De Regt 1999).

> Mach himself has ingeniously discussed the fact that no theory is absolutely true, and equally hardly any absolutely false either, but each must gradually be perfected, as organisms must according to Darwin's theory. By being strongly attacked, a theory can gradually shed inappropriate elements while the appropriate residue remains (Boltzmann 1974d, 153).

In summary, Boltzmann in effect identifies scientific truth with adequacy, and adequacy with weak correspondence between scientific theories and Nature. As noted above, unlike strong correspondence, by which a theory would, as it were, mirror Nature, his concept of weak correspondence implies theoretical pluralism, meaning: *(1)* no scientific theory is unique, since uniqueness implies dogmatism; *(2)* all theories are unavoidably incomplete, as a theory is an image of Nature, with completeness implying dogmatism; and *(3)* all scientific theories are temporally limited, that is, no theory is definitive: all theories will one day be replaced by better ones.

# 3. Theory assessment and the concept of reality

Boltzmann himself further developed various of the consequences of the views outlined above. As a confirmed Darwinist, he extended the notion of theoretical pluralism to reason itself. For Boltzmann, the brain was a device, an organ for creating images of the world, which, given how useful these images had proved to the conservation of the human species had reached a certain degree of perfection. The brain could be thought of as a physical structure ruled by evolution. So, reason itself would inevitably evolve by means of natural selection.

Another important point to emphasize is that although theories are representations and personal theoretical options can influence their formulation, they are *not* entirely arbitrary, thanks to the principle of weak correspondence between scientific theories and the external world – a principle that is implicit in Boltzmann's philosophical thinking. The basic aim of any theory is to represent phenomena that happen in Nature, and a successful theory can achieve this to a remarkable extent. Such a theory may use symbols or a specific mathematical language just as conventions. However, since Nature itself must be represented in any theory (corresponding weakly, that is, since weakly to it), then conventions will always be limited only to those aspects of the model that are not perceived, in that theory, as being directly dictated by Nature. Thus, theories cannot be seen as just conventions, because as they are crafted by scientists as representations of non-arbitrary, natural phenomena, they become attached to these phenomena and end up saying something about what is really going on in Nature.

Alongside its quality as a representation, there is another factor that might attract scientists towards a given model: its predictive ability. Once a certain theoretical prediction is confirmed, scientific knowledge about Nature increases quantitatively due to the principle of weak correspondence. Accurate prediction is also relevant because any prediction is formulated within the context of a specific theoretical framework. As such, if a model proves capable of predicting unknown phenomena, it demonstrates all its explanatory power, since it is capable not only of bringing forth elements that are already known, but also of going even further by demonstrating the existence of other, as yet missing elements needed to enable a deeper, better-organized understanding of Nature. It is important to bear in mind that one of the most important aims of science is to expand and organize our knowledge about Nature, which means that the importance of a theory also lies in its ability to contribute to this expansion and organization. A preference for such theories makes them more likely to be used and developed than others – even by incorporating useful elements from poorer theories – to the point that the distance between them may grow so great that it ceases to be worthwhile for researchers to continue working with the poorer representations.

It should be noted that theoretical pluralism does not necessarily imply competition among different theories, but it does often imply complementarity. Since all theories have some explanatory power and do not all refer to the same set of ideas and phenomena, they all end up saying something about the physical processes at play in Nature. Therefore, far from being a problem for our better understanding of Nature, the emergence of different theories for similar sets of physical phenomena is essential for it. And if these different theories have elements that contradict each other, observations and experimentation offer us the first – but not the only – means to decide which elements of the emergent theories are to be discarded.

The importance of orthodoxy in science is clearly an issue for Boltzmann. As the validation of theories and models takes time, a certain degree of conservatism towards new theories and models and skepticism towards new observations is understandable, since it is not reasonable to build a sound conceptual and experimental body of scientific knowledge when there is constant change in the concepts of science. Such skepticism shows the existence of critique in science. Orthodoxy plays the healthy role of preserving the scientific knowledge obtained on stable bases until new theories prove to have sufficient internal consistency and experimental validation.

However, when strong conservatism and orthodoxy becomes deeply rooted, a situation may arise that, if not effectively and successfully challenged, may lead the scientific community to avoid any kind of change to established ideas (Ribeiro and Videira 1998). In such an environment, established theories crystallize, becoming dogmatic, and scientific debate ceases to exist. In other words, dogmatism is an enemy of scientific progress and the adoption of theoretical pluralism could be a way of avoiding its negative consequences. Boltzmann was a strong and lucid defender of the atomic perspective, which at the time was facing a growing number of powerful opponents who considered the atomic picture of the world outdated and proposed a different view based entirely on energy. Boltzmann feared that such a representation could become dogmatic, which would inevitably lead to the stagnation of physics. Boltzmann believed that once theoretical pluralism was accepted and fully incorporated into research activities, it would prevent any theory from being excluded from the scientific community.

If our scientific theories are constructions then they are always incomplete, which means that there must be knowledge outside the scientific establishment. This may be called *affirmative knowledge* in the sense that it lacks internal connections produced by theoretical thought. Affirmative knowledge states that phenomena are as they are because they are experienced and observed as such. However, if further investigation is carried out this knowledge will begin to transmute into phenomenological science. Because any scientific discipline is always incomplete various types of affirmative knowledge exist alongside science.

Theoretical pluralism also has consequences for *ethics* among scientists. As they are representations or images of Nature, scientific theories are therefore human constructions, reflecting Nature without governing it. These images are built by generations of scientists and this line of thinking leads us inevitably to suppose that the relationship among scientists themselves is therefore pivotal in this construction process. The existence of a free flow of ideas is essential for this process, as are ongoing critique and stimuli as motivations for scientists. All this implies that certain rules of behavior, or ethics, are needed for the scientific endeavor to progress and avoid stagnation. Although our scientific theories do reflect Nature, they are, and will ever be, human constructs.

Boltzmann's views about reality mean that the concepts of "reality" and "real" should not be mistaken one for the other. The former consists of a set of mental pictures of Nature created by our brains, whereas the latter is Nature itself, the external world, whose ultimate knowledge is and will ever be unknowable. Nature therefore constitutes what is real and is thus outside our brains, while reality is created in the form of mental pictures in our brains by its interface with what is real. Reason can therefore be thought of as being the "*logical rules*"[11] that govern reality. But, since reality is made of mental pictures of Nature, which will inevitably change with time, we can only conclude that *reality and reason must evolve*.

# 4. Boltzmann's defense of realism according to a more liberal interpretative attitude

One issue touched upon above is why Boltzmann oscillated between the two opposing epistemological camps of his day, sometimes seemingly on the side of the descriptivists, or positivists, and other times apparently siding with the representationalists, or realists. In this respect it is important to see Boltzmann first and foremost as a pragmatic – and therefore flexible – physicist whose

11 Boltzmann did not believe in "logical rules of reason" in the strict sense, but only as a result of the evolutionary processes undergone by our brains.

thinking was guided by practical aims towards his main objective, namely, the progress of science. Therefore, it seems reasonable to assume that Boltzmann did not really care which of the two camps was "right": he could oscillate between opposing viewpoints provided science progressed. If each camp were able to produce useful scientific insights that allowed science to evolve, especially avoiding dogmatism, he was apparently at ease with this oscillation.[12]

Although Boltzmann was aware of the oscillations in his own opinions, apparently considering unimportant any "incoherence" provided he could employ the methods of science to satisfactorily solve his problems, the same could not apply to some of his contemporaries, such as Georg Helm and Wilhem Ostwald (Deltete 1999). In fact, the question as to how to establish a balanced relationship between realism and the thesis that scientific theories are representations remained obscure until the 1970s. It was only then that this problem started to attract the attention of philosophers and historians of science, who recognized that Boltzmann's philosophical "incoherence" ought to be taken seriously because it could offer a fruitful key towards a more realistic understanding of scientific practice.

Several interpreters of Boltzmann's thinking from more recent decades, such as Cercignani (1998), Blackmore (1995, 1999), Bouverese (1991), Hiebert (1980), d'Agostino (1990), de Courtenay (2018), De Regt (1996, 1999, 2005), and Videira (1992, 1995), to cite a few, agree that Boltzmann's thinking is rich and sophisticated, despite lacking a degree of internal conceptual cohesion. In the words of de Courtenay (2018, 4), "the internal architecture of the physicist's philosophical thinking, however, remains opaque" (own translation). Nevertheless, it should be stressed, once again, that Boltzmann was not worried by this absence of clarity because he was suspicious as to the ability of philosophy to be of use to professional scientists (Videira 1995, 6, 12). His arguments in favor of realism ought therefore to ultimately be based on scientific achievements, stemming directly from what scientists themselves attain when they are realists.

It should be acknowledged, nonetheless, that Boltzmann's relative disinterest in refining his own epistemological reasoning made it hard for his ideas to be understood by readers and interpreters interested in his reflections – a state of affairs that remains to the present day. For this reason, it is perhaps worth returning to his writings to demonstrate more directly the realistic beliefs he advocated and their inherent sophistication. The Italian physicist Carlo Cercignani did so; indeed, his book on Boltzmann (Cercignani 1998) could be

12 We are grateful to a referee for pointing this out.

seen as a philosophical and scientific biography of the Austrian physicist. In it, he writes:

> Professional philosophers may find Boltzmann's ideas somewhat naive, but this is not the case. It is true that Boltzmann is a realist, but not a naive one. A philosopher reading the following pages [13] should always take into consideration that Boltzmann's vision of the world is the picture that every physicist has in mind when investigating his problems, even when he does not admit it (Cercignani 1998, 170).

This statement is valid, but it can be made even clearer by pointing out that Boltzmann was not only interested in defending the centrality of the image (*Bild* or *Vorstellung*) in scientific practice, but also in analyzing how this image relates to the external world, a topic that has been discussed by others.

> The difficulties encountered in defining Boltzmann's position in the debate on atomism, and more broadly on realism, are largely due to his conception of theories as "images" of phenomena. According to this conception, "the task of theory is to construct an image (*Abbild*) of the external world" by accomplishing "with greater magnitude what happens on a small scale each time we form an idea (*Vorstellung*) (de Courtenay 2018, 3; own translation).

It is our interpretation that Boltzmann never abandoned his views on realism when dealing with scientific problems, especially when it came to the kinetic theory of gases. This was the case even when he recognized that in scientific problems (De Regt 1996, 1999) a better characterization was needed of the way in which realism could coexist *amicably* with the belief that scientific theories are images, since purporting that theories and models were copies of natural phenomena was untenable.

De Regt (2005) seems to agree with Cercignani's interpretation above, albeit without making reference to him, but rather than just affirming Boltzmann's credentials as a realist, and he qualifies Boltzmann's realism.

> A few opponents of atomism, most notably Ernst Mach, rejected scientific realism completely, even the metaphysical thesis. But most physicists were less radical. James Clerk Maxwell and Ludwig Boltzmann defended atomism and the kinetic theory by developing sophisticated realist positions, particularly by qualifying the epistemic thesis without rejecting it completely. Their ideas on metaphysics, epistemology, and methodology were shaped in the context of the debates about the value of the kinetic theory, and in turn influenced their further work on this theory (...) The case study shows a mutual influence between science and philosophy, and illustrates that scientists can be 'flexible realists': they are sometimes willing to adapt their realist views to the situation at hand (De Regt 2005, 206).

Among the various interesting points in the citation above, it is worth emphasizing that Boltzmann drew on changes in the scientific context, like the difficulties faced by the kinetic theory of gases, to refine his position on realism. The kinetic theory of gases was one that Boltzmann held dear, but if science indicated that epistemic or ontological changes were required in his hypotheses, models, and theories, Boltzmann saw no reason to cling on to them.

[13] Here Cercignani refers to pages on his own book (Cercignani 1998, chap. 10, 170-197).

> Yet, Boltzmann remained committed to a realistic view of the kinetic theory and atomism in general; it was only with respect to specific molecular models that he weakened his realistic goals: as we have seen in the previous section, he admitted that his dumbbell model does not faithfully represent real diatomic molecules in all respects (De Regt 2005, 221).

In accepting that his "dumbbell models" did not faithfully represent the real world, Boltzmann not only reasserted his faith in realism, but also, as pointed out by De Regt, turned back to another cherished viewpoint of his, namely, that concepts are creations of human beings. Moreover, these creations are indispensable because without them the task of doing science would become a chimera. Even if there were freedom in the production of scientific concepts, these ought to be formulated as if they were instructions that permit acceptable and viable interventions in the world.

# 5. Conclusions

This paper builds upon the perspective advanced by Ribeiro and Videira (1998) on Boltzmann's epistemological ideas on realism to discuss theory assessment. The genesis of Boltzmann's thinking was introduced in its original context, followed by a discussion on how they are related to scientific truth and theoretical assessment. The argumentation is based on the concept of theoretical pluralism, which is a consequence of scientific theories being viewed as representations, or images, of Nature. This is Boltzmann's key epistemological conclusion, meaning that the external world can be represented by different theories, none being absolutely true or absolutely false since they all say something about the external world. The unfolding discussion reached at the viewpoint that truth is defined by the adequacy of a theory and must always be provisional, meaning that there could never be any ultimate theories since all will one day be replaced by better ones.

Ribeiro and Videira (1998) focused on Boltzmann's defense that scientific theories are only representations and, therefore, could not tell how Nature really is. It also focused on his concerns on how to avoid dogmatism, which if not challenged would slow down, or even block, the evolution of science. Here we add that in the final years of his life Boltzmann realized that although such a reasoning was valid and correct, it was also insufficient. As shown in Section 4 the discussion had to move to analyzing the relationship between the image and the real, to realism, if dogmatism were to be avoided.

Although we cannot know whether there will ever be a consensus on the adequate interpretation of the philosophical ideas developed by Boltzmann, it seems to us that it is impossible to deny that he developed his ideas with the intention of using them to counteract the creation of scientific and philosophical dogmas. Boltzmann believed in the importance of his ideas, mainly because they could help strengthen and improve science, philosophy, and human life in general. His ideas could serve as inspiration for the eternal path taken by the human species in its struggle for survival and should not be regarded as the last word on such topics. For Boltzmann, genuine progress would be achieved if human beings could continuously improve themselves. His definition of certainty frames this perspective: "We design as true the actions whose results produce desired things, as well as the representations that direct us to act that way" (Boltzmann 1905b, 164; our translation). In other words, Boltzmann seems to place "action," "intention," "representation," and "truth" on the same plane as importance. These four concepts constituted a balanced and indivisible web, in his view, and the key idea responsible for maintaining this balance was his concept that scientific theories are representations.

It follows that our theoretical constructions, or images of Nature, can be applied to physics itself. If physics is a construction, it follows that physics must also follow Darwin's natural selection to solve its conceptual and foundational issues. In this sense, Darwinism could be applied to science in general. As such, it is erroneous to seek explanations beyond human activity and beyond what human beings can attain through the ordinary senses and their brains.

The activity of constructing images of the external world and the process of honing them to become more adequate representations are instinctive to human beings (Boltzmann 1905b, 179), which means they are not the sole privilege of scientists. Ever since their appearance, humans have been using their ability to construct internal mental pictures of the external world to interact with their surroundings and with other living beings. In order to survive, humans need to act and intervene in the environment. All representations, including those we call scientific theories, have a practical side that allows humankind to make use of Nature to its own benefit. Therefore, representing is a human activity that is as ordinary as any other, and there must be a common substrate to any human activity. Thus, representing means acting, since it requires practical decisions and choices to solve the problems humanity encounter.

The philosophical reflections of the Austrian physicist in the final years of his life went beyond the strict limits of epistemology, since he realized that the reality analyzed by science should be in itself analyzed to make sense of his pluralism. Boltzmann concluded that reality as discussed by scientific theories is not different from the reality to which humans are immersed. Human beings reach reality by means of their actions, whatever they are, and those include the scientific practice. Representing and acting can then be understood as forming a web.

## ACKNOWLEDGEMENTS

We are grateful to the reviewers and the managing editor Tara Mahfoud for various useful comments, suggestions and recommendations which led to the improvement of this paper. Antonio A. P. Videira is thankful for the support provided by the Centro Brasileiro de Pesquisas Físicas.

## FUNDING

Antonio A. P. Videira acknowledges the financial support of the Brazilian federal funding agency CNPq, grant number 306612/2018-6 (level 1C), as well as the Prociência Program of the Universidade do Estado do Rio de Janeiro.

## DISCLOSURE STATEMENT

The authors report no potential conflict of interests.